# Peer review vs bibliometrics: which method better predicts the scholarly impact of publications?[1]

**Authors:** Giovanni Abramo[1]*, Ciriaco Andrea D'Angelo[2], Emanuela Reale[3]

**Affiliations:**

[1]Laboratory for Studies in Research Evaluation, Institute for System Analysis and Computer Science (IASI-CNR). National Research Council, Rome, Italy

[2]University of Rome "Tor Vergata", Dept of Engineering and Management, Rome, Italy

[3]Research Institute on Sustainable Economic Growth (IRCRES-CNR). National Research Council, Rome, Italy

**Abstract**
In this work, we try to answer the question of which method, peer review vs bibliometrics, better predicts the future overall scholarly impact of scientific publications. We measure the agreement between peer review evaluations of Web of Science indexed publications submitted to the first Italian research assessment exercise and long-term citations of the same publications. We do the same for an early citation-based indicator. We find that the latter shows stronger predictive power, i.e., it more reliably predicts late citations in all the disciplinary areas examined, and for any citation time window starting one year after publication.

**Keywords**
*Research evaluation, scientometrics, publication quality, scientific advancement.*

# 1. Introduction

Research evaluation is a fundamental activity in any science system. It is functional to continuous improvement and informs research policy and management.

Scientists collect and analyze prior knowledge encoded in verbal and written forms and add value to it—producing new knowledge, which they nearly always try to transfer by encoding it in written form. The main aim is to make it accessible to other scientists, thus contributing to further scientific and/or technical advancements. Journal and book editors, conference organizing committees, etc. then use a peer review system to evaluate submitted manuscripts and decide which are worth publishing. Research agencies assess research proposals and proponents to allocate competitive funding, while research institutions evaluate the research performance of their staff and the potential of candidates for research grants or positions within their organization. Lastly, national governments assess the performance of domestic research institutions.

Whichever the entity under evaluation, the basic unit of analysis is generally the publication, i.e. the knowledge produced and encoded therein in written form by its authors. Among others, the aim of evaluation is to predict the future impact of publications[2] on further scientific and/or technical advancements. We use the term *predict*, because assessing the actual impact of a publication would require sufficient time for the knowledge embedded in it to impact all possible future scientific and/or technical advancements. Policy makers and managers, who hope to make informed decisions, cannot afford to wait long enough for the publication lifecycle to be completed so as to conduct evaluation. Therefore, they have to deal with the embedded trade-off between level of accuracy and timeliness in impact measurement.

The question posed by scholars and practitioners in the field is whether the impact of publications can be better evaluated by human judgment or through the use of bibliometric indicators (when available and reliable), or by drawing on both (informed peer review). The answer very much depends on the purposes that the evaluation has to serve and on its context (scale, budget, and time constraints).

The aim of this empirical work is to contribute to making better-informed decisions in relation to the above choice. We compare the peer review evaluation scores and early citation-based scores of a large set of publications to a benchmark, which represents a robust proxy of total impact of such publications. In this way, we can determine which method shows better predictive power, by discipline and length of citation time window. Our research is based on the first Three-Year Research Assessment Exercise (VTR 2001-2003) carried out in Italy, which was entirely peer reviewed and for which the evaluation scores assigned by reviewers to each publication submitted for evaluation are available.

The findings of this study can support decision makers, who might wish to replace the costly and time-consuming peer reviewing process with cheaper and faster bibliometric analyses in large-scale research assessment, or integrate it with bibliometric scores in small-scale assessment.

The rest of the paper is organized as follows. Section 2 deals with the different perspectives feeding the scientific debate on the two evaluation approaches. Section 3 illustrates the key features of the VTR, while Section 4 sets out our methods and data. Section 5 describes and discusses the results of the comparison, while the final section, Section 6, presents our conclusions.

---

[2] In this work, we deal with the evaluation of research output encoded in written form and indexed in bibliographic repertories (such as Scopus and Web of Science, WoS), i.e. after fellow specialists have assessed its suitability for publication.

**2. The peer review vs bibliometrics debate**

The decision to use citation-based indicators as a proxy for the impact of scientific production is founded on phenomena that have been widely studied by sociologists of science. In a narrative review of studies on the citing behavior of scientists, Bornmann and Daniel (2008) analyze the motivations that lead scholars to cite the works of others. Their findings show that "citing behavior is not motivated solely by the wish to acknowledge intellectual and cognitive influences of colleague scientists, since the individual studies reveal also other, in part non-scientific, factors that play a part in the decision to cite." Nevertheless, "there is evidence that the different motivations of citers are not so different or randomly given to such an extent that the phenomenon of citation would lose its role as a reliable measure of impact."

In the sociology of science, two different theories of citing behavior have been developed: the normative theory and the social constructivist view. The first, based on the work of Robert Merton (1973), states that, through citations, scientists acknowledge that they are indebted to colleagues whose results they have used, meaning that citations represent intellectual or cognitive influence on scientific work.

The social constructivist view on citing behavior is based on the constructivist theory in the sociology of science (Latour, 1987; Cetina, 1981). This approach contests the assumptions at the basis of the normative theory and thus weakens the validity of evaluative citation analysis. Constructivists argue that "scientific knowledge is socially constructed through the manipulation of political and financial resources and the use of rhetorical devices" (Cetina, 1981), meaning that citations are not linked in any direct and consequential manner to the scientific content of the cited articles. The bibliometric approach relies instead on the assumption that this link is strong and direct, meaning that citational analysis can be the main instrument to evaluate the scientific impact of research publications.

The alternative option for assessing the impact of a publication is through human judgment. Indeed, it is frequently held that the basis for research evaluation lies in experts reviewing the work of their colleagues. Nevertheless, peer evaluation is not without weaknesses, as it is clearly susceptible to certain built-in distortions due to subjectivity (Moxham & Anderson, 1992; Horrobin, 1990). These can arise at several levels, including when peer judgment is used to gauge output quality, but also in the earlier step of selecting the experts that will carry out the assessment. In fact, the exceptionally specialized nature of present-day research makes it difficult to identify the most appropriate experts (Abramo, D'Angelo, & Viel, 2013) and, even when they accept to serve as reviewers, their ability to express fair judgments is by no means a given. The rapidity of scientific advances can also pose serious difficulties in contextualizing the quality of research output produced a number of years earlier. Would a reviewer truly be able to disregard all subsequent scientific advances when expressing a judgment?

Furthermore, peer review evaluations can be affected by real or potential conflicts of interest, such as the tendency to give more positive evaluations to the output of famous scholars than to that of younger, less established researchers, or the failure to recognize all qualitative aspects of the output (which increases with the increasing specialization and sophistication of the work being assessed). In addition, in the peer review methodology, mechanisms to assign merit are established autonomously by the various evaluation panels and/or individual reviewers, thus exposing assessment to internal disagreement and comparisons based on this methodology to potential distortions. "Bias in peer review, whether intentional or inadvertent, is widely recognized as a confounding factor in efforts to judge the quality of research" (Pendlebury, 2009). Evidence from the Italian research assessment exercise VQR 2004-2010 suggests that internal peer review agreement is quite low (Bertocchi, Gambardella, Jappelli, Nappi, & Peracchi, 2015). Reale and Zinilli (2017) solicited more research efforts and new rules aimed at improving accountability and reliability of peer review, shedding light on how it contributes to innovative research paths.

On the other hand, evaluative bibliometrics is also fraught with a number of limitations. For instance, it is not possible to apply citation indicators to the entire range of research outputs, but only to publications indexed in bibliographic repertories. Yet, the most frequent objection to the use of citation counts in evaluating research output is whether and to what extent citations are a certification of merit. It has however been shown that, while negative citations may occur, these are actually rare and do not disrupt large-scale analyses (Pendlebury, 2009).

Citation-based indicators can also be affected by certain forms of manipulation, such as excessive recourse to self-references and cross citations, or citations of articles in the same journal but not related to the content of the citing publication (under pressure by editors eager to increase the impact indicator of their journals). In response to the tendency to overexpose the drawbacks of self-citations, Pichappan and Sarasvady (2002) list nine reasons for author self-referencing, and conclude that "they cannot be rejected *in toto*, as they have a complex nature and require careful interpretations."

Another factor often mentioned as criticism of citation analyses is the phenomenon of "delayed recognition", also in the case of more mature works (Garfield, 1980; van Raan, 2004; Ke, Ferrara, Radicchi, & Flammini, 2015). Other objections concern article reviews, which are seen as inflating citations, and the "once highly cited, always highly cited" phenomenon. Finally, it must be noted that evaluative citation-based analysis is unable to capture impact outside the scientific system, such as on practitioners (e.g. a physician applying a new pharmacological protocol after reading relevant literature), in education (e.g. scientists serving in the transmission of new knowledge to students), and on technology (e.g. scientists acting as consultants to industries and governments). That is why the impact measured by evaluative bibliometrics is more precisely referred to as scholarly impact (Abramo, 2018).

Glänzel (2008) analyzes the validity of what he calls "seven bibliometrics myths", with cognitive and methodological background. Here are the conclusions of his analysis: "Although there is always a grain of truth in bibliometrics myths too, the generality of their statements is disproved on the bases of methodological studies and by referring to typical counterexamples. It is shown how and where the logical fallacy lies in the inference from the reality behind the myths leading to the erroneous generalization of the actual statements." On the contrary, Herrmannova, Patton, Knoth, and Stahl (2018) draw attention to the fact that widely used research metrics poorly distinguish research that strongly influences later developments from works that predominantly discuss the current state of affairs.

What emerges from the literature is that the pros and cons of the peer review and metric approaches in the evaluation of publications appear to be balanced, making it difficult to establish which would be preferable: the context variables and evaluation goals are likely to shift the weight in favor of one or the other.

Although no consensus has been reached on which method ought to be preferred, it is possible to ascertain whether the two approaches lead to similar evaluations. A number of studies have actually delved into the correlation between the results of the two methodologies.

Indeed, a positive correlation has been detected in the evaluation of individual research products (Bertocchi, Gambardella, Jappelli, Nappi, & Peracchi, 2015; ANVUR, 2013; Bornmann & Leydesdorff, 2013; Allen, Jones, Dolby, Lynn, & Walport, 2009; Reale, Barbara, & Costantini, 2007; van Raan, 2006; Aksnes & Taxt 2004; Oppenheim & Norris 2003; Rinia, van Leeuwen, van Vuren, & van Raan, 1998; Oppenheim, 1997), individual researchers (Vieira & Gomes, 2018; Vieira, Cabral, & Gomes, 2014a; Vieira, Cabral, & Gomes, 2014b; Cabezas-Clavijo, Robinson-García, Escabias, & Jiménez-Contreras, 2013; Bornmann, 2011; Meho & Sonnenwald, 2000), as well as research institutions (Pride & Knoth, 2018; Abramo, D'Angelo & Di Costa, 2011; Franceschet & Costantini, 2011; Thomas & Watkins, 1998).

Yet, many of these studies display limitations: they analyze the correlation looking at a small number of scientific sectors, they adopt bibliometric indicators that are not always the most appropriate, and/or their inferential analyses are incorrect. For instance, concerning the latest Italian research assessment exercise, VQR 2011-2014, Abramo and D'Angelo (2016) reveal that the bibliometric indicator used to predict the future impact of publications, a weighted combination of citation and IF percentiles (Ancaiani et al., 2015), is not valid. As a consequence, the studies by Bertocchi, Gambardella, Jappelli, Nappi and Peracchi (2015) and Alfò, Benedetto, Malgarini, and Scipione (2017) on peer review vs bibliometrics correlation (which use the above indicator) are affected by the same problem. Moreover, Baccini, Barabesi, and De Nicolao (2018) point out that the samples on which the two studies above are founded cannot be considered "representative" of the population of articles submitted to the VQR.

Other works have come to opposite conclusions. Differently from previous analyses on the Research Excellence Framework (REF) in the UK (Mryglod, Kenna, Holovatch, & Berche, 2015; Mahdi, D'Este, & Neely, 2008; Thomas & Watkins, 1998; Taylor, 2011), the steering group for the review of the role of metrics in research assessment found that, in the case of the 2014 REF, individual metrics yielded significantly different outcomes from the peer review process, showing that metrics cannot provide a like-for-like replacement for REF peer review (Wilsdon et al., 2015). Traag and Waltman (2019) object that, in the REF context, proper comparisons between metrics and peer review should be made at the institutional level, rather than at the level of individual publications. They conclude that for some disciplines—namely Clinical Medicine, Physics, and Public Health, Health Services & Primary Care—agreement between metrics and peer review is similar to internal peer review agreement.

The conclusion by Wilsdon et al. (2015) is also reached by Baccini and De Nicolao (2016), who challenge the data and findings of ANVUR (2013) on the agreement between peer review and bibliometrics in the evaluation of products within the second Italian research assessment exercise, VQR 2004-2010. The authors explain that "the degree of agreement has to be interpreted, for all research fields, as unacceptable, poor or, in a few cases, as, at most, fair."

Regardless of similarities in their results, none of these analyses resolve the question of which methodology, peer review vs bibliometrics, would be more accurate and reliable, based on the citation time window considered, in predicting the long-term scholarly impact of publications. The remainder of this paper will illustrate the approach adopted to shed light on this matter and the results emerged from our research.

**3. The first Italian research assessment exercise, 2001-2003 VTR**

To better understand the methodology (presented in the next section) used to answer our research question, let us first describe the 2001-2003 VTR. The aim of the VTR was to assess R&D performed by universities and public research institutions. Each institution evaluated (102 in total; 64,000 researchers) was asked to submit its 2001-2003 research products. The number of products could not exceed 50% of the full-time-equivalent (FTE) research staff[3], and the typologies admitted were limited to articles, books and chapters of books, proceedings of national and international congresses, patents, designs, performances, exhibitions, artifacts, and works of art. Hence, the VTR excluded purely editorial activities, texts and software for teaching purposes, congress abstracts, trials and routine analyses, and internal technical reports.

Twenty disciplinary panels made up of high-level peers (151 panelists, 79 of which from Italian universities, 37 from abroad, 19 from domestic research institutions, and 16 from the industry) were appointed to assess a total of 17,329 research products (72% articles, 23% books and book chapters, 2% patents, and 3% miscellaneous), with additional support from external experts (2 experts for each product evaluation at least; 6,661 in total) (Cuccurullo, 2006). At the end of the peer review process, each product was given a final judgment expressed on a four-point rating scale: Excellent (E) = 1 if the product quality fell within the top 20% of international standards; Good (G) = 0.8 if it was between 80% and 60%; Acceptable (A) = 0.6 if it was between 60% and 40%; and Limited (L) = 0.2 if it was below 40%.

For every university and research institution, the sum of the scores for each evaluated product was divided by the total number of products submitted (P). So the rating of generic organization *i* was calculated according to the following formula:

$$Rating_i = \frac{1}{P_i}\left[\sum_i E_i + 0.8 \sum_i G_i + 0.6 \sum_i A_i + 0.2 \sum_i L_i\right]$$

[1]

**4. Methods and data**

**4.1 Publication quality vs impact: What do we want to measure?**

Traditionally, it has been held that the proper basis for research evaluation consists in experts reviewing the work of their colleagues. No surprise then that the comparison between bibliometrics and peer assessment has been widely acknowledged as a way to validate citation impact metrics (Garfield 1979; Bornmann & Daniel, 2005; Harnad,

---
[3] To account for time devoted to teaching activities, 1 professor equals 0.5 FTE.

2008; Kreiman & Maunsell, 2011; Kulczycki, Korzeń, & Korytkowski, 2017, Traag & Waltman, 2019).

When comparing the results of peer review and bibliometrics assessments, an implicit assumption is that both approaches evaluate the same attributes of a publication. Leaving all the limits and caveats of bibliometrics aside, the normative theory of citing behavior holds that citation-based indicators measure the scholarly impact of publications. Peers, instead, tend to assess the "quality" of publications. The question then is whether quality equals scholarly impact. But what is quality in the first place? According to Martin and Irvine (1983), "quality is a property of the publication and the research described in it. It describes how well the research has been done, whether it is free from obvious "error", how aesthetically pleasing the mathematical formulations are, how original the conclusions are, and so on". Further, quality "is not just intrinsic to the research, but is something judged by others who, with differing research interests and social and political goals, may not place the same estimates on the quality of a given paper. Even the same individual may evaluate the quality of a paper differently at different times because of progress in scientific knowledge and shifts in his or her location." Then, as the authors themselves state, quality is highly subjective.

In addition to Martin and Irvine (1983)[4], various other scholars have defined quality and distinguished it from impact (Cole & Cole, 1973; Mingers & Leydesdorff, 2015; Sugimoto & Larivière, 2018, p. 66). Further references can be found in the review work by Leydesdorff, Bornmann, Comins, and Milojević (2016), which summarizes the perspectives of a number of scholars on the difference between quality and impact. In most cases, it emerges that quality definitions are simply stated without any supporting arguments.

Abramo (2018) questions whether a distinction between "quality" and "impact" is truly necessary: "Contrary to what most believe quality does not need to be a subjective attribute. According to the ISO 9000: 2015 International Standard, which provides the fundamental concepts, principles and vocabulary for quality management systems (QMS) and the foundation for other QMS standards, the technical definition of quality is "the degree to which a set of inherent characteristics fulfils a requirement[5]." Given that in general the ultimate requirement of a publication is that it provides impact on future scientific advancement, quality needs to refer to impact, and the measure of quality and impact would then be synonymous." In their recent review of the literature, Aksnes, Langfeldt, and Wouters (2019) argue that "citations reflect aspects related to scientific impact and relevance, although with important limitations. On the contrary, there is no evidence that citations reflect other key dimensions of research quality."

The equivalence between quality and impact is thus an open question that divides the scientific community, and proposing a definitive answer is beyond the scope of this paper. Regardless of individual opinions, it is hard to deny that scholarly impact represents one of the fundamental merits of published research output. This is because, although quality and impact might not be equivalent, a thorough assessment of the merits of a publication must necessarily take scholarly impact into account, or else end up relying solely on formal scientific rigor.

---

[4] More precisely, Martin and Irvine (1983) argue that "it is necessary to distinguish between, not two, but three concepts - the "quality", "importance", and "impact" of the research described in a paper", whereby "importance" of a publication refers to its *potential* influence on the advance of scientific knowledge.

[5] ISO (the International Organization for Standardization) is a worldwide federation of national standards bodies. ISO 9000: 2015 is accessible at https://www.iso.org/obp/ui/#iso:std:iso:9000:ed-4:v1:en, last accessed 12 June, 2019.

The comparison between peer review and bibliometrics proposed here focuses on the ability to predict the long-term scholarly impact of a publication. However, since evaluation generally occurs before the publication life-cycle is completed, the question is whether expert judgment might turn out to be better at predicting long-term impact than early citation-based indicators. Thus, we do not question the idea that late citation counts, i.e. scholarly impact, serve as a benchmark for determining the preferable approach (and its predictive power).

**4.2 Methods**

The research products submitted to the VTR were produced between 2001 and 2003. To carry out a bibliometric assessment of their impact, the first step is to put together the dataset, matching the evaluated products to those indexed in a bibliographic repository containing their citations. The bibliographic repository used here is WoS.

Step two is to count late citations (January 2018) for each publication in the dataset. Late citation counting (minimum citation time window: 14 yrs; maximum: 17 yrs) can be considered a good proxy for long-term impact (Abramo, Cicero, & D'Angelo, 2011), and represents a reliable benchmark to compare the ability of peer review evaluation vis-à-vis early-citation-based evaluation to predict long-term scholarly impact.

Step three consists in classifying publications by WoS subject category (SC) and disciplinary area (DA).

Step four concerns measuring each research product's early citation based impact, respectively one year, two years, and three years after publication. The analysis does not extend beyond a three-year citation window, because decision makers generally require timely assessments. To measure the scholarly impact, a weighted combination of normalized citations and impact factor (IF) of the hosting journal is applied to each publication. Weights differ across citation time windows and SCs.

Having defined normalized IF and citations of publication *i*, respectively as:

$$x^i = \frac{IF_k^i}{\overline{IF}_k}$$

[2]

$$y_t^i = \frac{c_t^i}{\bar{c}_t}$$

[3]

where:
- *t* is the citation time window, with $1 \leq t \leq 3$;
- $c_t^i$ is the number of citations received by publication i, *t* years after publication;
- $\bar{c}_t$ is the average number of citations received *t* years after publication by all cited publications of the same year and SC of publication *i*;
- *k* is the publication year, with $2001 \leq k \leq 2003$;
- $IF_k^i$ is the impact factor of the journal hosting publication *i*, at publication year;
- $\overline{IF}_k$ is the average of impact factors of all journals falling in the SC of publication *i*, at publication year.

The scholarly impact *SI* of publication *i*, published in year *k*, is the following:

$$SI_k^i = b_0^t + b_1^t x^i + b_2^t y_t^i \qquad [4]$$

We use the parameters *b* which best predict future impact, as calculated by Abramo, D'Angelo, and Felici (2019).

In step five, we repeat step four for each 2001-2003 world publication in the same SC as those in the dataset. The reason for this is to be able to measure the percentile rank of each publication in the dataset, per citation time window and SC. As we will see in step six, the percentile rank allows us to compare bibliometric and peer review scores. WoS bibliometric indicators (citations and IF) for all world publications in the years and SCs included in the dataset analyzed are retrieved from the University of Leiden CWTS WOSKB database, the CWTS version of Clarivate Analytics Web of Science database.[6]

Step six involves classifying each publication included in the dataset into the VTR classes, according to the relevant bibliometric percentile rank:
- Percentile rank ≥ 80 → Excellent (E) = 1
- 60 ≤ Percentile rank < 80 → Good (G) = 0.8
- 40 ≤ Percentile rank < 60 → Acceptable (A) = 0.6
- Percentile rank < 40 → Limited (L) = 0.2

At this point, we have comparable (early and late) bibliometric and peer review scores. We can then compare the predictive power of peer review vs bibliometric evaluation as a function of the citation time window and SC.

The agreement between the benchmark evaluation and, respectively, peer review and bibliometric assessments is measured through Cohen's kappa agreement test and Lin's concordance correlation coefficient.

We draw our comparison analyzing the dataset as a whole, as well as each DA and citation time window. This stratification is valuable, for instance, when making operational decisions regarding large-scale national assessment exercises. Indeed, it might emerge that peer review evaluation is more reliable than the bibliometric approach for products in DA *X* but not for those in DA *Y*, or we might see that after a certain time *t* (i.e. after a given citation time window) bibliometrics is always more reliable. This makes it possible to identify the most effective methodology depending on the age and field of the research product in question.

**4.3 Data**

The VTR-evaluated products made available to us belong to eight panels, corresponding to the following DAs: 1 - Mathematics and computer science; 2 - Physics; 3 - Chemistry; 4 - Earth sciences; 5 - Biology; 7 - Agricultural and veterinary sciences; 8 - Civil engineering; 9 - Industrial and information engineering. These represent all Sciences but Medicine. 27,771 professors, equal to 45.8% of the total teaching staff of Italian universities, worked in the DAs under observation during the three-year period analyzed (2001-2003).

---
[6] We take this opportunity to thank the CWTS for making its database accessible to us for research purposes.

A total of 9,225 research products were submitted to the above panels. In some cases, co-authors from different institutions submitted the same products. 8,086 were individual publications indexed in WoS. A few products showed a publication date different from 2001-2003; others had no IF; others were sent to different panels (by different co-authors) and received different evaluation scores. After excluding all the above cases, the final dataset amounted to 7,276 publications, as shown in Table 1. This corresponds to 11.8% of the WoS-indexed scientific production of all Italian academics, ranging from a minimum of 7.2% in Industrial and information engineering to a maximum of 27.3% in Earth sciences.

*Table 1: Number of publications in the dataset, per year and disciplinary area*

| Disciplinary area | 2001 | 2002 | 2003 | Total dataset (a) | Total population** (b) | a/b |
|---|---|---|---|---|---|---|
| 1 - Mathematics and computer science | 170 | 276 | 274 | 720 | 6,258 | 11.5% |
| 2 - Physics | 488 | 534 | 519 | 1,541 | 12,414 | 12.4% |
| 3 - Chemistry | 255 | 361 | 384 | 1,000 | 12,958 | 7.7% |
| 4 - Earth sciences | 165 | 202 | 214 | 581 | 2,126 | 27.3% |
| 5 - Biology | 462 | 499 | 541 | 1,502 | 14,545 | 10.3% |
| 7 - Agricultural and veterinary sciences | 192 | 222 | 248 | 662 | 4,053 | 16.3% |
| 8 - Civil engineering | 107 | 115 | 119 | 341 | 2,323 | 14.7% |
| 9 - Industrial and information engineering | 283 | 337 | 370 | 990 | 13,706 | 7.2% |
| Total* | 2,106 | 2,521 | 2,649 | 7,276 | 61,592 | 11.8% |

*\* The figures on the last line do not match the column total due to products falling into multiple DAs being counted more than once.*
*\*\* Total 2001-2003 WoS publications authored by Italian professors in each DA.*

## 5. Results

In presenting our results, we refer to the bibliometric score (BS) and VTR peer review score (PR) of each publication. Both scores can be 0.2, 0.6, 0.8, or 1. Depending on the citation time window, we have four BS scores for each publication: BS_1Y (2Y, 3Y), for citations measured 1 (2, 3) years after publication, and BS_2018 for citations measured in 2018, representing a proxy for total impact, or benchmark.

In order to determine the extent of the agreement between the predictive scores (PR, BS_1Y, BS_2Y, and BS_3Y) and the benchmark (BS_2018), we use Cohen's kappa coefficient $k$ (Cohen, 1960), a statistic which measures inter-rater agreement for qualitative (categorical) items (Sheskin, 2003), defined as:

$$k = \frac{p_0 - p_e}{1 - p_e}$$

[5]

where $p_0$ is the observed proportion of agreement between the raters, and $p_e$ is the proportion of agreement expected by chance. The upper limit, k = 1, is reached when the two raters are in perfect agreement, while k ≤ 0 means that the observed agreement is less than or equal to the agreement expected by chance.

Since more than two ordered nominal categories are in place (the four VTR classes E, G, A, L), we use the weighted kappa, $k_w$, (Cohen, 1968), where the weights indicate the seriousness of the disagreement. The most common weights are linear and quadratic weights (Fleiss, Levin, & Myunghee, 2003). In the case at hand, each pair of scores is regarded as two raters expressing different judgments on the same publication, resulting in the assignment of each article to one of the four merit classes. We choose to apply linear weights (0; 0.33; 0.67; 1), where the lowest possible disagreement, i.e. any one-merit class disagreement (disagreements between E and G, G and A, A and L), is weighted one third (0.33) of the highest possible disagreement (1), i.e. the disagreement between E and L. According to Fleiss, Levin, and Myunghee (2003), the agreement can be considered:

- Poor, if $k_w \leq 0.40$
- From fair to good, if $0.40 < k_w \leq 0.75$
- Excellent, if $k_w > 0.75$.

The results of the analysis are reported in Table 2. All $k_w$ values are statistically significant (last column of Table 2) but, in terms of practical significance, an absolute comparison of $k_w$ values unequivocally indicates that the level of agreement between the benchmark and PR is poor, and much lower than between the benchmark and BS, also when considering a one-year citation window (BS_1Y). If the timeframe is extended to a two-year citation window (BS_2Y), the value of $k_w$ (0.4698) falls within the second category identified by Fleiss, Levin, and Myunghee (2003), so that the agreement is "fair to good". To further improve the predictive power of the bibliometric indicator and achieve excellent agreement, a larger citation time window is needed (Abramo, Cicero, & D'Angelo, 2011; Abramo, D'Angelo, & Felici, 2019).

*Table 2: Cohen's $k_w$ test for agreement between predictive scores (PR, BS_1Y, BS_2Y, and BS_3Y) and the benchmark score (BS_2018) of publications in the dataset*

|       | Agreement | Expected agreement | Cohen's $k_w$ | Standard error | Z     | Prob>Z |
|-------|-----------|--------------------|----------------|----------------|-------|--------|
| PR    | 75.08%    | 69.92%             | 0.1716         | 0.0082         | 20.93 | 0.000  |
| BS_1Y | 77.12%    | 64.74%             | 0.3512         | 0.0084         | 41.58 | 0.000  |
| BS_2Y | 81.70%    | 65.48%             | 0.4698         | 0.0085         | 55.42 | 0.000  |
| BS_3Y | 84.70%    | 66.30%             | 0.5460         | 0.0085         | 63.92 | 0.000  |

To shed further light on the level of agreement with the benchmark, Table 3 presents the cross-classification of the data concerning the scores of PR, BS_1Y, and BS_3Y compared to the benchmark score, BS_2018. The sum of the values along the main diagonal (shaded area) of each matrix represents the number of cases in which the pairs of raters are in perfect agreement. If the PR scores are considered, this is true for 40.4% of publications (23+289+1,125+1,504 out of 7,276). Perfect agreement between raters is found in 48.4% of the cases when using BS_1Y, while the figure increases to 60% if we look at BS_3Y scores. As for the merit classes, peer review displays a higher number of cases of perfect agreement than bibliometrics for publications with "Good" impact, while this number is lower for all the other classes.

The sum of the values in the cells above the diagonal represents the number of cases in which either peer review or bibliometric predictions overestimate the long-term impact of publications. On the contrary, the sum of the values below the diagonal refers to cases of underestimation. The data in Table 3[7] reveal that the PR score overestimates the benchmark value in 26.8% of the cases (121+210+73+710+245+589 out of 7,276) and underestimates it in 32.8% of the cases (51+59+415+41+362+1,459 out of 7,276). Moreover, the BS_1Y score overestimates the benchmark value in 24.6% of the cases and underestimates it in 26.9% of the cases. Lastly, the cases of underestimation and overestimation are both equal to 20.0% of the total for BS_3Y.

---

[7] For reasons of space, we omit reporting the case of BS_2Y score because, as expected, results are just between those for BS_1Y score and BS_3Y score.

The cases in which the greatest overestimation is found are:
- 73 for PR (PR=1, BS_2018=0.2)
- 17 for BS_1Y (BS_1Y=1, BS_2018=0.2)
- 2 for BS_3Y (BS_3Y=1, BS_2018=0.2)

Conversely, the greatest underestimation occurs in:
- 41 cases with PR=0.2 and BS_2018=1
- 182 cases with BS_1Y=0.2 and BS_2018=1
- 36 cases with BS_3Y=0.2 and BS_2018=1

We observe that as compared to bibliometrics, peer-review tends to significantly overestimate low impact publications. There could be different explanations for that. We exclude upfront the possible distortion induced by the authors' affiliation with prestigious universities, simply because, with the exception of three tiny advanced schools, the rest of the universities in the sciences show little variability in research performance. A long-standing non-competitive higher education system has not given birth to elite universities in Italy (Abramo, Cicero, & D'Angelo, 2012).

After empirical testing, we observe instead that the reputation of the journals could have contributed to the overestimation of lower impact publications, in fact the higher the relevant journal IF the higher the overestimation of the impact of the publications by the reviewers.

Another co-determinant might be the social proximity (ties of kinship, belonging to the same institution, advisory roles, etc.) between reviewers and authors, whose distorting effect has already been demonstrated in the context of evaluation committees of candidates for academic career advancement (Abramo, D'Angelo, & Rosati, 2015).

Results further show that the predictive power of bibliometrics for short citation time windows is particularly weak for publications with low citations (182 cases of underestimation, with BS_1Y=0.2 and BS_2018=1), weaker than peer-review (41 cases of underestimation).

*Table 3: Contingency table of predictive scores (PR, BS_1Y, BS_3Y) and the benchmark score (BS_2018) of publications in the dataset*

|  |  | PR | | | | BS_1Y | | | | BS_3Y | | | | Total |
|---|---|---|---|---|---|---|---|---|---|---|---|---|---|---|
|  |  | .2 | .6 | .8 | 1 | .2 | .6 | .8 | 1 | .2 | .6 | .8 | 1 |  |
| BS_2018 | .2 | 23 | 121 | 210 | 73 | 212 | 105 | 93 | 17 | 285 | 113 | 27 | 2 | *427* |
|  | .6 | 51 | 289 | 710 | 245 | 316 | 299 | 451 | 229 | 269 | 458 | 478 | 90 | *1,295* |
|  | .8 | 59 | 415 | 1,125 | 589 | 294 | 331 | 665 | 898 | 138 | 373 | 934 | 743 | *2,188* |
|  | 1 | 41 | 362 | 1,459 | 1,504 | 182 | 227 | 610 | 2,347 | 36 | 101 | 537 | 2,692 | *3,366* |
| *Total* |  | *174* | *1,187* | *3,504* | *2,411* | *1,004* | *962* | *1,819* | *3,491* | *728* | *1,045* | *1,976* | *3,527* | *7,276* |

In order to validate the results of Cohen's $k_w$ test, we use Lin's concordance correlation coefficient ($\rho_c$), which indicates how well a given measurement compares to a "gold standard" measurement (Lin, 1989, 2000). Like a correlation, $\rho_c$ ranges from -1 to 1, with perfect concordance at 1. It cannot exceed the absolute value of $\rho$, Pearson's correlation coefficient.

The data in Table 4 confirm a greater degree of concordance between the bibliometric scores and the benchmark than between the peer review scores and the benchmark. Specifically, the $\rho_c$ value of BS_1Y is twice that of PR. In the case of BS_3Y, the $\rho_c$ value is equal to 0.667, pointing to moderate concordance between the predictor and the benchmark (McBride, 2005).

*Table 4: Concordance correlation coefficients between predictive scores (PR, BS_1Y, BS_2Y, and BS_3Y) and the benchmark score (BS_2018) of publications in the dataset*

|       | Lin's $\rho_c$ | Std Err. | [95% CI]    | P     | Pearson's $\rho$ |
|-------|----------------|----------|-------------|-------|------------------|
| PR    | 0.221          | 0.011    | 0.199-0.242 | 0.000 | 0.228            |
| BS_1Y | 0.436          | 0.009    | 0.418-0.454 | 0.000 | 0.453            |
| BS_2Y | 0.576          | 0.008    | 0.561-0.591 | 0.000 | 0.591            |
| BS_3Y | 0.667          | 0.006    | 0.654-0.679 | 0.000 | 0.675            |

We repeat the analysis at the level of single DAs in order to ascertain whether, and to what extent, the agreement between the predictors and the benchmark varies across scientific domains. This segmentation is particularly important, considering that the rapidity with which citations accrue varies across disciplines, and the predictive power of bibliometrics with short citation time windows is weak in the case of lowly cited publications. Table 5 displays the values of Cohen's $k_w$ test, which shows that in Civil engineering, Industrial and information engineering, Agricultural and veterinary sciences; and Mathematics and computer science, the predictive power of bibliometrics is lower than in the rest of the disciplines, just because of the above said reason.

Results invariably confirm, anyway, the evidence emerged from the overall analysis. The bibliometric scores are always in greater agreement with the benchmark, compared to peer assessment, even when considering a short timeframe, i.e. a one-year citation window. In Agricultural and veterinary sciences (DA 7), the agreement between PR and BS_2018 is close to zero (Cohen's $k_w$ = 0.077), while that between BS_1Y and BS_2018 is equal to 0.2774. In Physics, the strongest concordance is found between PR and BS_2018 (Cohen's $k_w$ = 0.2508), but the value remains noticeably lower than that concerning the bibliometric scores.

The bibliometric predictor for the three-year citation window (BS_3Y) shows fair to good agreement with the benchmark in all DAs. The Cohen's $k_w$ is never below 0.45, with a peak of 0.6522 in Biology. Indeed, in Biology, fair to good agreement between the bibliometric predictor and the benchmark is already present when the one-year citation window is analyzed (Cohen's $k_w$ above 0.4).

Table 6 reports the results of the analysis performed using Lin's concordance correlation coefficient, which perfectly overlap the values of Cohen's $k_w$. Lin's $\rho_c$ between PR and BS_2018 is always below 0.3, and in half of the DAs it is below 0.2. Conversely, the concordance of the BS_1Y predictor is twice that of PR in practically all DAs, with the sole exception of Civil engineering, where it is still higher. If we look at the three-year citation window, the bibliometric predictor (BS_3Y) displays good Lin's concordance correlation, never below 0.55 and with a peak of 0.774 in Biology— a DA with a high Lin's $\rho_c$ (equal to 0.5) already one year after publication.

*Table 5: Cohen's $k_w$ agreement coefficient between predictive scores and the benchmark score of publications in the dataset, by disciplinary area (standard errors in brackets)*

| DA* | PR | BS_1Y | BS_2Y | BS_3Y |
|---|---|---|---|---|
| 1 | 0.1394 (0.025) | 0.2838 (0.024) | 0.3984 (0.025) | 0.4806 (0.026) |
| 2 | 0.2508 (0.019) | 0.3665 (0.018) | 0.4676 (0.018) | 0.5188 (0.018) |
| 3 | 0.1308 (0.022) | 0.2988 (0.023) | 0.4432 (0.023) | 0.5354 (0.024) |
| 4 | 0.1311 (0.029) | 0.2904 (0.027) | 0.4364 (0.028) | 0.5362 (0.029) |
| 5 | 0.1779 (0.018) | 0.4079 (0.019) | 0.5714 (0.019) | 0.6522 (0.019) |
| 7 | 0.0771 (0.023) | 0.2774 (0.0250) | 0.3756 (0.025) | 0.4560 (0.025) |
| 8 | 0.1597 (0.036) | 0.2459 (0.030) | 0.3586 (0.031) | 0.4533 (0.034) |
| 9 | 0.1084 (0.021) | 0.2711 (0.021) | 0.4050 (0.022) | 0.4930 (0.022) |

*\* 1 - Mathematics and computer science; 2 - Physics; 3 - Chemistry; 4 - Earth sciences; 5 - Biology; 7 - Agricultural and veterinary sciences; 8 - Civil engineering; 9 - Industrial and information engineering*

*Table 6: Lin's concordance correlation coefficient between predictive scores and the benchmark score of publications in the dataset, by disciplinary area (standard errors in brackets)*

| DA* | PR | BS_1Y | BS_2Y | BS_3Y |
|---|---|---|---|---|
| 1 | 0.171 (0.033) | 0.377 (0.028) | 0.494 (0.025) | 0.597 (0.022) |
| 2 | 0.283 (0.022) | 0.481 (0.019) | 0.625 (0.015) | 0.670 (0.013) |
| 3 | 0.180 (0.029) | 0.399 (0.026) | 0.583 (0.021) | 0.685 (0.017) |
| 4 | 0.200 (0.038) | 0.398 (0.030) | 0.537 (0.027) | 0.662 (0.022) |
| 5 | 0.252 (0.023) | 0.499 (0.018) | 0.686 (0.013) | 0.774 (0.010) |
| 7 | 0.153 (0.034) | 0.328 (0.030) | 0.458 (0.027) | 0.549 (0.024) |
| 8 | 0.205 (0.050) | 0.313 (0.035) | 0.455 (0.034) | 0.576 (0.031) |
| 9 | 0.139 (0.028) | 0.334 (0.025) | 0.509 (0.022) | 0.626 (0.019) |

*\* 1 - Mathematics and computer science; 2 - Physics; 3 - Chemistry; 4 - Earth sciences; 5 - Biology; 7 - Agricultural and veterinary sciences; 8 - Civil engineering; 9 - Industrial and information engineering*

## 4. Conclusions

The evaluation of scientific publications is a crucial activity and constitutes the basis for assessing research performance at several different levels, i.e. individual, institutional, regional, etc. The debate around the pros of the peer review method vs bibliometrics to evaluate research products has taken on increasing relevance and it is far from showing any sign of converging opinions. Scholars and practitioners tend to prefer one approach over the other, even though the informed peer review method is gaining growing support.

So far, most studies have focused on analyzing to what extent the results yielded by the two methodologies are aligned, usually for the purpose of validating the bibliometric approach (which is both faster and cheaper) as a viable substitute for peer review in large-scale research assessment exercises. The underlying axiom, made more or less explicit, is that the proper basis for research evaluation consists in experts reviewing the work of their colleagues.

There is still no consensus on whether peer review and bibliometrics assess the same attributes of publications, but impact on future scientific advances is undoubtedly a fundamental dimension of evaluation. In this research, we have compared the two methodologies along that very dimension, more specifically looking at their ability to predict the scholarly impact of publications, without losing sight of the shortcomings and limitations affecting both methods.

We have ascertained that, in the sciences under examination (which exclude medicine), the predictive power of bibliometrics proves superior to that of peer review as early as one year after publication, and this is true for all DAs. Furthermore, the longer the citation time window, the more accurate the bibliometric approach becomes in predicting future impact.

Thus, we can conclude that the real question is no longer whether either methodology is better at assessing the scholarly impact of a given research product, but rather what sorts of trade-offs exist between accuracy and timeliness of research evaluation or, in other words, what decision makers want to evaluate and how long they are willing to wait in order to make accurate, informed decisions.

Having determined that bibliometrics is superior to peer review in predicting scholarly impact, further studies may be aimed at investigating how to strengthen such predictive power in order to improve the accuracy of the assessments and relevant policy decisions. The fact that this papers analyzes a specific case within the Italian setting, implying potentially problematic issues related to the poor quality of the reviewers' work, should not invalidate the generalizability of our findings, also because the assessment panels in our case study included several foreign scientists and experts from a broad range of institutions.

Our results might be of interest to those who establish research assessment procedures at various levels, with the purpose of examining the scientific production of the subjects under evaluation. This is particularly true for those who formulate large-scale research assessment exercises, in which the choice of the peer review approach inevitably implies a reduction in the amount of scientific production assessed, for time and cost reasons, and consequent additional distortions in the scores and rankings of research products (Abramo, Cicero, & D'Angelo, 2013). In view of the results illustrated here, decision makers may come to more informed decisions on whether to use peer review, bibliometrics, or both, depending on the objectives and on the desired level of accuracy, timeliness, and costs of the assessment.